# Memory-Assisted Universal Source Coding

Ahmad Beirami and Faramarz Fekri
School of Electrical and Computer Engineering
Georgia Institute of Technology, Atlanta GA 30332, USA
Email: {beirami, fekri}@ece.gatech.edu

The problem of the universal compression of a sequence from a library of several small to moderate length sequences from similar context arises in many practical scenarios, such as the compression of the storage data and the Internet traffic. In such scenarios, it is often required to compress and decompress every sequence individually. However, the universal compression of the individual sequences suffers from significant redundancy overhead [1], [2]. In this paper, we aim at answering whether or not having a memory unit in the middle can result in a fundamental gain in the universal compression. We present the problem setup in the most basic scenario consisting of a server node $S$, a relay node $R$ (i.e., the memory unit), and a client node $C$, as depicted in Fig 1. We assume that server $S$ wishes to send the sequence $x^n$ to the client $C$ who has never had any prior communication with the server, and hence, is not capable of memorization of the source context. However, $R$ has previously communicated with $S$ to forward previous sequences from $S$ to the clients other than $C$ (not shown in Fig. 1), and thus, $R$ has memorized a context $y^m$ shared with $S$. Note that if the relay node was absent the source could possibly apply universal compression to $x^n$ and transmit to $C$ whereas the presence of memorized context at $R$ can possibly reduce the communication overhead in $S$-$R$ link.

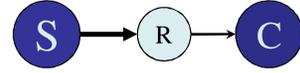

Fig. 1. The basic memory-assisted compression scenario.

We let the information source located in $S$ be parametric with $d$-dimensional parameter vector $\theta$ that is unknown a priori. Let $l_n$ denote the universal length function for the compression of the sequence $x^n$. Denote $l_{n|m}$ as the universal length function for $x^n$ with context memorization, where the encoder (at $S$) and the decoder (at $R$) have access to a memorized context $y^m$. Our goal is to investigate the fundamental gain of the context memorization in the memory-assisted universal compression of the sequence $x^n$ over conventional universal source coding. To do so, we compare two schemes:

- Ucomp (Universal compression of an individual sequence with no context memorization), in which a sole universal compression is applied on the sequence $x^n$ without context memorization.
- UcompM (Universal compression of an individual sequences with context memorization), in which the encoder and the decoder will both have access to the memorized context $y^m$ from the same source, and they utilize $y^m$ as the context for the compression of the sequence $x^n$.

Let $H_n(\theta)$ be the source entropy given $\theta$. Further, let $Q(l_n, l_{n|m}, \theta) \triangleq \frac{\mathbf{E} l_n(X^n)}{\mathbf{E} l_{n|m}(X^n)}$, where $\mathbf{E}$ denotes expectation with respect to the true unknown parameter $\theta$. Let $\epsilon$ be such that $0 < \epsilon < 1$. We denote $g(n, m, \epsilon)$ as the fundamental gain of the context memorization (relative to the conventional universal compression without memory) for a sequence of length $n$ from family $\mathcal{P}^d$ of parametric sources using a memory of length $m$ for a fraction $1 - \epsilon$ of the sources. That is

$$g(n, m, \epsilon) \triangleq \sup_{z \in \mathbb{R}} \left\{ z : \ \mathbf{P}\left[Q(l_n, l_{n|m}, \theta) \geq z\right] \geq 1 - \epsilon \right\}.$$

The following theorem presents a lower bound on the gain of memory-assisted source coding.

**Theorem 1** *Assume that the parameter vector $\theta$ follows Jeffreys' prior in the universal compression of the family of parametric sources $\mathcal{P}^d$. Then,*

$$g(n, m, \epsilon) \geq 1 + \frac{\bar{R}_n + \log(\epsilon) - \hat{R}(n,m)}{H_n(\theta) + \hat{R}(n,m)} + O\left(\frac{1}{n\sqrt{m}}\right),$$

where $\bar{R}_n$ is the average minimax redundancy, and $\hat{R}(n,m) \triangleq \frac{d}{2}\log\left(1 + \frac{n}{m}\right) + 2$.

Our result shows that UcompM achieves more than $50\%$ memorization gain as compared to Ucomp in the compression of a first-order Markov sequence of length $n = 128$kB with a memory of size $m = 8$MB. Further, with sufficient memory, $x^n$ can be compressed to the entropy rate.

This material is based upon work supported by the National Science Foundation under Grant No. CNS-1017234.